\documentclass[11pt,a4paper]{article}

\usepackage{authblk}
\usepackage{fullpage}
\usepackage[round]{natbib}
\usepackage{url}
\usepackage{color, colortbl}
\usepackage{float}
\usepackage{graphicx}

\DeclareGraphicsExtensions{.png,.jpg}

\newcommand{\keywords}[1]{\small{\textbf{Keywords:} {#1}}}

\usepackage{fancyhdr}
\fancypagestyle{plain}{%
  \fancyhf{}%
  \fancyhead[L]{
  \begin{minipage}{0.5\textwidth}
  \tiny{
\noindent Preprint of an article published in Journal of Integrative Neuroscience, Online Ready, pp. 1-12, 2017.\\
\noindent http://www.worldscientific.com/worldscinet/jin }
\end{minipage}}
	\fancyhead[R]{
  \begin{minipage}{0.5\textwidth}
	\begin{flushright}
  	\tiny{
		\noindent \textcopyright copyright World Scientific Publishing Company\\		
		\noindent doi: 10.1142/S0219635216500266
	}
	\end{flushright}
   \end{minipage}}

}%

\title{\huge\bfseries Pupil size behavior during on line processing of sentences}

\author[1]{G. Fern{\'a}ndez\footnote{Gerardo Fernández and Juan Biondi contributed equally to this work.}}
\author[1,2]{J. Biondi$^{*}$}
\author[2]{S. Castro}
\author[1,3]{O. Agamennoni}

\affil[1]{Universidad Nacional del Sur (UNS), Instituto de Investigaciones en Ingenier{\'i}a El{\'e}ctrica (IIIE) (UNS-CONICET), Bah{\'i}a Blanca, Buenos Aires, Argentina.}

\affil[2]{Universidad Nacional del Sur (UNS), Departamento de Ciencias e Ingenier{\'i}a de la Computaci{\'o}n, Laboratorio de visualizaci{\'o}n y computaci{\'o}n gr{\'a}fica (VyGLab), Bah{\'i}a Blanca, Buenos Aires, Argentina.}

\affil[3]{Comisi{\'o}n de Investigaciones Cient{\'i}ficas de la Provincia de Buenos Aires (CIC), Argentina}

\date{}

\begin{document}

\definecolor{LightGray}{gray}{0.9}



\vspace{-35pt}       
       
\maketitle


\vspace{-35pt}

\begin{abstract}
In the present work we analyzed the pupil size behavior of forty subjects while they read well defined sentences with different contextual predictability (i.e., regular sentences and proverbs). In general, pupil size increased when reading regular sentences, but when readers realized that they were reading proverbs their pupils strongly increase until finishing proverbs' reading. Our results suggest that an increased pupil size is not limited to cognitive load (i.e., relative difficulty in processing) because when participants accurately recognized words during reading proverbs, theirs pupil size increased too. Our results show that pupil size dynamics may be a reliable measure to investigate the cognitive processes involved in sentence processing and memory functioning.
\end{abstract}


\section{Introduction}
The relationship between cognitive load (i.e. total amount of mental effort) and pupil dilation have attracted the attention of many researchers (See \citet{laeng2012pupillometry} for an overview).  Although these studies differ in how cognitive load is implemented, they showed that there exist a strong direct correlation between cognitive load and pupil size. 
In a non-linguistic context, \citet{preuschoff2011pupil} concluded that pupil size (and therefore, presumably, cognitive load) increases when a stimulus is less expected. \citet{preuschoff2011pupil} studied pupil dilation while participants performed a simple gambling task. He found that pupil size correlated with unexpectedness, not with the gambling outcome itself (i.e. surprise causes pupil dilation). 

Whether unexpectedness of words in sentences also results in pupil dilation is still an open question. When reading a sentence, each current word is integrated with the past words and predictions about upcoming words are generated (e.g., \citet{just1982paradigms, fernandez2014eye, fernandez2015diagnosis}). The amount of cognitive effort required to process a given word reflects the interplay of word processing and expectancy driven processes \citep{rayner1998eye, kliegl2006tracking}. \citet{esterman2009perceptual} showed that contextual hints facilitate the information processing. \citet{gazzaley2013top} concluded that working memory performance in improved by contextual hints. The relationship between low predictable upcoming words and cognitive load has been observed in reading studies; the time needed to be read is directly correlated with its surprising values, which is related to different sentence comprehension phenomena’s, like garden-path effect \citep{brouwer2010modeling} and anti-locality effects \citep{levy2008expectation,frank2012early}. In reading, gaze time duration at each word is correlated with surprising and with low predictable upcoming words (e.g., \citet{boston2008parsing, demberg2008data, smith2008optimal, fernandez2014eye}). 

Nevertheless, there are only a few works in psycholinguistics analyzing pupil's behavior during reading sentences. \citet{engelhardt2010pupillometry} showed that, in comparison with the matching situation, a mismatch between the syntactic and prosodic structure of auditorily presented sentences increase pupil size. \citet{piquado2010pupillometry} found a pupil response to syntactic complexity and sentence length in a sentence-listening study.

To the best of our knowledge, there exist only a few published studies in which pupillometry is applied during reading well-defined sentences. \citet{raisig2012role} developed a study with written descriptions of simple events in everyday activities. When the order of presentation was incongruent with the actual temporal order of the described activities they observed an increase in pupil dilation. \citet{just1982paradigms} evaluated reading time and pupil size in a study were they compared object and subject-relative clauses. They found increased reading times and pupil dilatation on the object-relatives that are known to be more difficult to process \citep{hakes1976understanding}. Moreover, a semantically implausible word increased pupil size compared to a plausible-word. Conversely, \citet{papesh2012memory}, encounter little perceptual resistance when processing the same words spoken by different speakers, each of whom has a unique vocal structure, speaking rate and pattern of intonation. Further, participants showed that encoding effort of processing spoken words is related to subsequent memory strength. This effect was not limited to encoding, since when participants recognized old items during test, their pupils were again more dilated.
In the present study, we examined pupil behavior during on-line sentences processing.
We examined whether word properties embedded in sentences with different contextual predictabilities (i.e., regular sentences and proverbs) affect pupillary responses. Using proverbs as reading material allows to examine whether a semantic context facilitates reading processes \citep{katz2001moment, fernandez2014eye, fernandez2015diagnosis}. Then, it would be possible to counter check whether two kinds of stimulus, an easier one (proverbs) and a less facilitated (regular sentences), affect pupil responses.

\section{Methods}
\subsection{Participants}
The group of readers consisted of forty persons (mean age: 58, $SD=4:1$) with professional qualification and no evidence of cognitive decline or impairment in activities of academic and daily living. 
The mean education trajectory of the readers was 16.1 years ($SD=1:0$). To assess whether subjects comprehended the texts, they were presented with a three alternative multiple-choice question about the sentence in progress on $20\%$ of the sentence trials. 
Participants answered the questions by moving a mouse and choosing the response with a mouse click. Overall mean accuracy was $96\%$ ($SD=3:1\%$).
For a sentence corpus description see \citet{fernandez2014eye}. 
For Apparatus technical specifications and for eye movement data analyses see \citet{fernandez2015diagnosis}.

\subsection{Analytical strategy for pupil data}
We concentrated our analysis on the period that begins with the presentation of the first word of the sentence and ends with the last word. 
Also, a pupil size normalization procedure was applied on each individual trial, dividing pupil size data by the mean baseline value, defined as a period equivalent to $25\%$ of the normalized time before sentence presentation.

\subsection{Statistical analysis}
Our analyses are based on linear mixed models (LMMs). 
We used the \textit{lmer} function of the \textit{lme4} package (version 0.999999-2) \citep{bates2013linear} for estimating fixed and random coefficients.
This package is supplied in the R system for statistical computing under the GNU General Public License (Version 2, June 1991). 
The dependent variable was the normalized pupil diameter. 
Fixed effects in LMM terminology correspond to regression coefficients in standard linear regression models. They can be used to estimate slopes or differences between conditions. The following fixed effects were entered into the model: logit predictability, log frequency, 1/word length. We specified contrasts of sentence type (proverbs vs. regular sentences). In addition, we allowed for varying slopes of pupil diameter with participants and items (sentences) by setting random slopes for participants and items. Instead of estimating differences between conditions, random effects estimate the variance that is associated with the levels of a certain factor. For the LMMs we report regression coefficients (bs), standard errors (SEs), and t-values ($t=b/SE$). Our criterion for referring to an effect as signifficant is $t = b/SE > 1:96$.

\section{Results}

\begin{table}[htpb]
\centering
\caption{Parameter estimates for fixed effects of Linear Mixed Models. Threshold of significance is set at $t=±1.96$.}
\begin{tabular}{ l c c c }
	 &  & \textbf{Pupil size} &  \\ 
	 & \textbf{M} & \textbf{SE} & \textbf{t-value} \\ \rowcolor{LightGray}
	\textbf{Fixed effects} &  &  &  \\ 
	Normalized Pupil diameter & 0.988 & 0.050 & \textbf{19.49} \\ 
	Word Number & 0.000 & 0.001 & 0.20 \\ 
	Predictabilities (logit) & -0.013 & 0.003 & \textbf{-3.62} \\ 
	Frequencies (log) & -0.005 & 0.002 & \textbf{-2.61} \\ 
	1/Length (characters) & -0.077 & 0.032 & \textbf{-2.37} \\ \rowcolor{LightGray}
	\textbf{Sentence type} &  &  &  \\ 
	Proverbs vs Regular sent. & 0.024 & 0.012 & \textbf{1.99} \\ 
	Proverbs vs Regular sent. x Word Number & 0.003 & 0.000 & \textbf{-2.46} \\
	Proverbs vs Regular sent. x Predictabilities (logit) & 0.015 & 0.005 & \textbf{3.03} \\ 
	Proverbs vs Regular sent. x Frequencies (log) & 0.002 & 0.003 & 1.38 \\ 
	Proverbs vs Regular sent. x Length (characters) & -0.010 & 0.016 & -0.61 \\ \rowcolor{LightGray}
	\textbf{Variance components} &  & \textbf{Variance} & \textbf{SD} \\
	Sentence (n=140) &  & 0.000 & 0.017 \\
	Subject (n=40) &  & 0.093 & 0.305 \\ 
	Residual (n=11215) &  & 0.006 & 0.079 \\ 
\end{tabular}
\label{tab:estimates}
\end{table}

\begin{figure}[H]
\centering
\includegraphics[width=0.7\columnwidth]{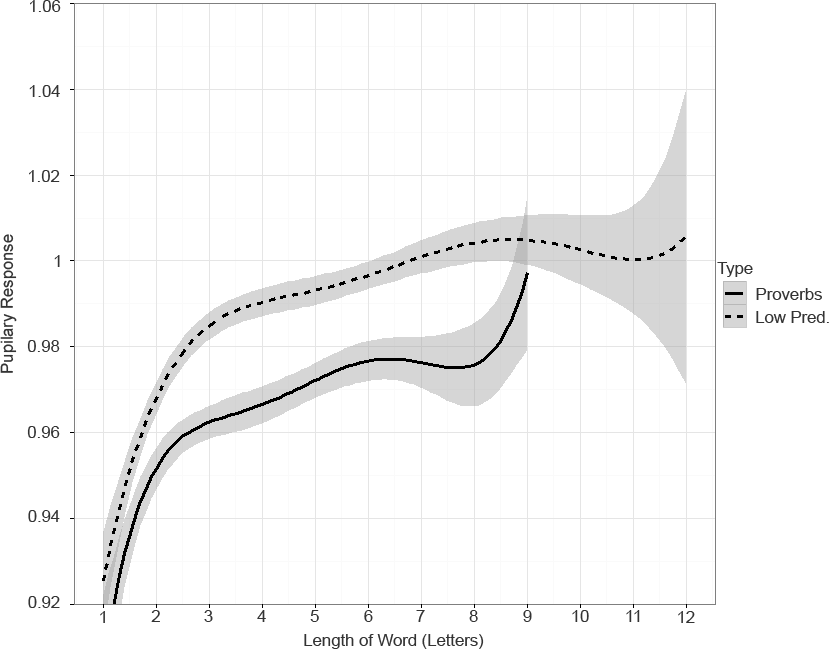}
\caption{Effect of the length of word on normalized pupil dilatation, broken down by regular sentences and proverbs. Panel reflects regression of normalized pupil dilatation on word on respective length. Shaded areas are $95\%$ confidence intervals.}
\label{fig:length}
\end{figure}

\begin{figure}[H]
\centering
\includegraphics[width=0.7\columnwidth]{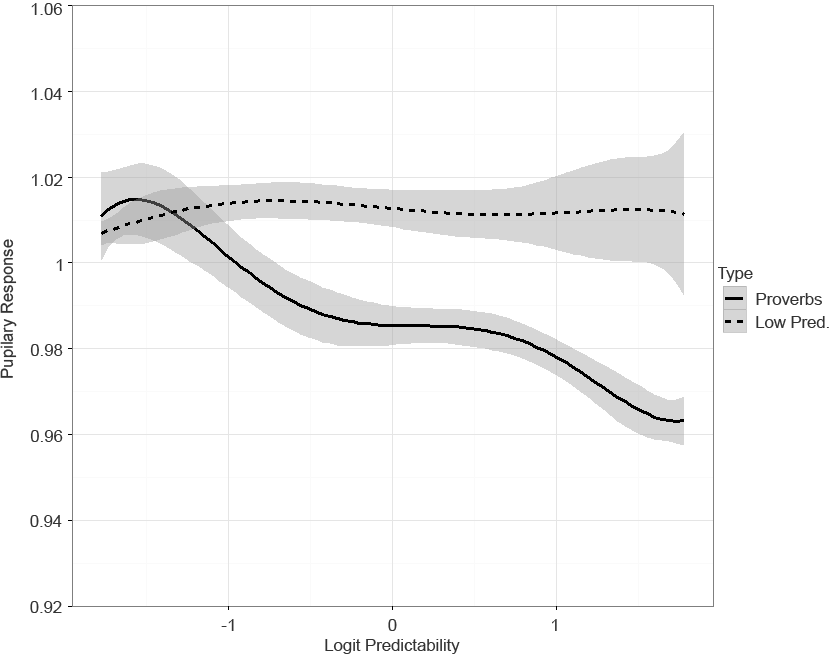}
\caption{Effect of the predictability of word on normalized pupil dilatation, broken down by regular sentences and proverbs. Panel reflects regression of normalized pupil dilatation on word on respective logits of predictability. Shaded areas are $95\%$ confidence intervals.}
\label{fig:pred}
\end{figure}

\begin{figure}[H]
\centering
\includegraphics[width=0.7\columnwidth]{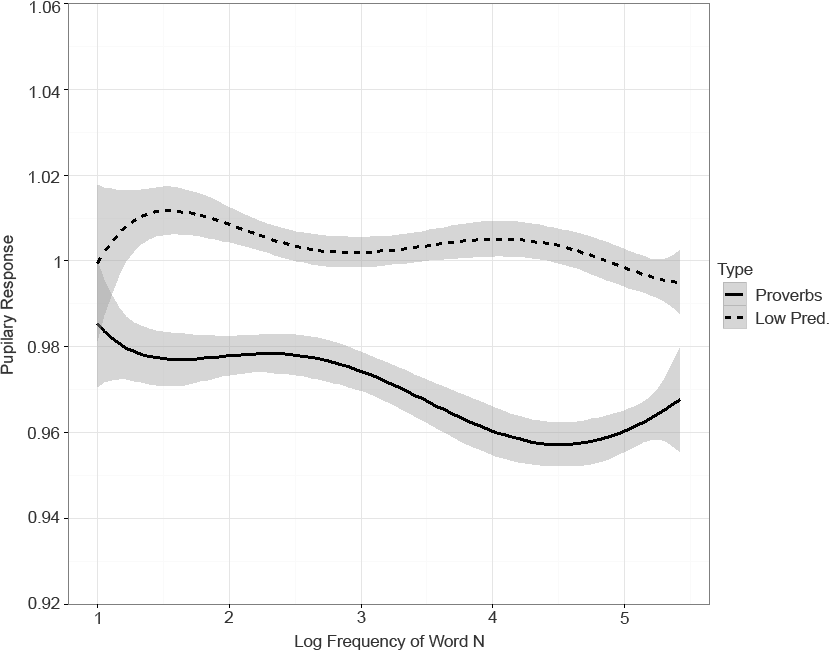}
\caption{Effect of the frequency of word on normalized pupil dilatation, broken down by regular sentences and proverbs. Panel reflects regression of normalized pupil dilatation on word on respective log of frequency. Shaded areas are $95\%$ confidence intervals.}
\label{fig:freq}
\end{figure}

\begin{figure}[H]
\centering
\includegraphics[width=0.7\columnwidth]{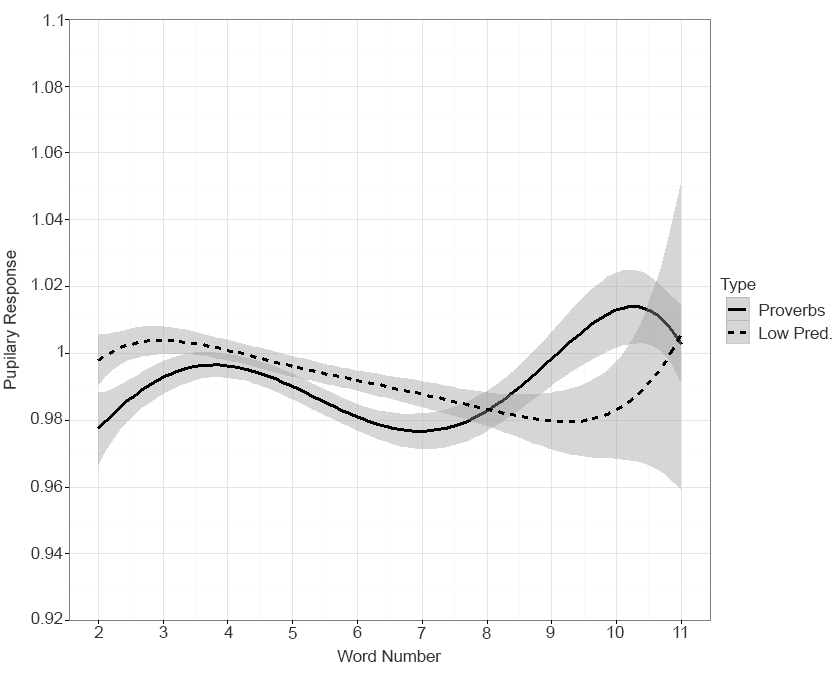}
\caption{Effect of number of word in sentences on normalized pupil dilatation, broken down by regular sentences and proverbs. Panel reflects regression of normalized pupil dilatation on word on respective number of word. Shaded areas are $95\%$ confidence intervals.}
\label{fig:nw}
\end{figure}

In Table \ref{tab:estimates} we report, (a) main effects when averaging over all predictors i.e., collapsing sentence type (proverbs vs. regular sentences); (b) interactions of word properties, and word number x sentence type. 
The normalized pupil diameter as a function of sentence type, and word properties are displayed in Figures \ref{fig:length}, \ref{fig:pred}, \ref{fig:freq} and \ref{fig:nw}. 
The values in Figures are partial effects (for an example of this technique, see \citet{hohenstein2014semantic}).

As shown in Table \ref{tab:estimates}, the mean pupil diameter significantly increased when reading regular sentences ($t=1.99$) compared to proverbs, it seems that regular sentences increased readers' cognitive load. 
When we evaluated whether word properties -averaging over all predictors- affected pupil responses we observed how 1/word length, word frequency and word predictability exerted a significant effect on pupil dilatation ($t=-2.37$, $t=-2.61$, $t=-3.62$, respectively). 
As shown in Figure \ref{fig:length}, longer words increased pupil dilatation because probably longer words are more difficult to process. 
Quite the contrary, both more frequent and more predictable words decreased the pupil diameter. 
Word number is not affecting significantly the pupil size when considering averaging collapsing all predictors ($t=0.20$) (See Table \ref{tab:estimates} and Figure \ref{fig:nw}).

\begin{figure}[H]
\centering
\includegraphics[width=1\columnwidth]{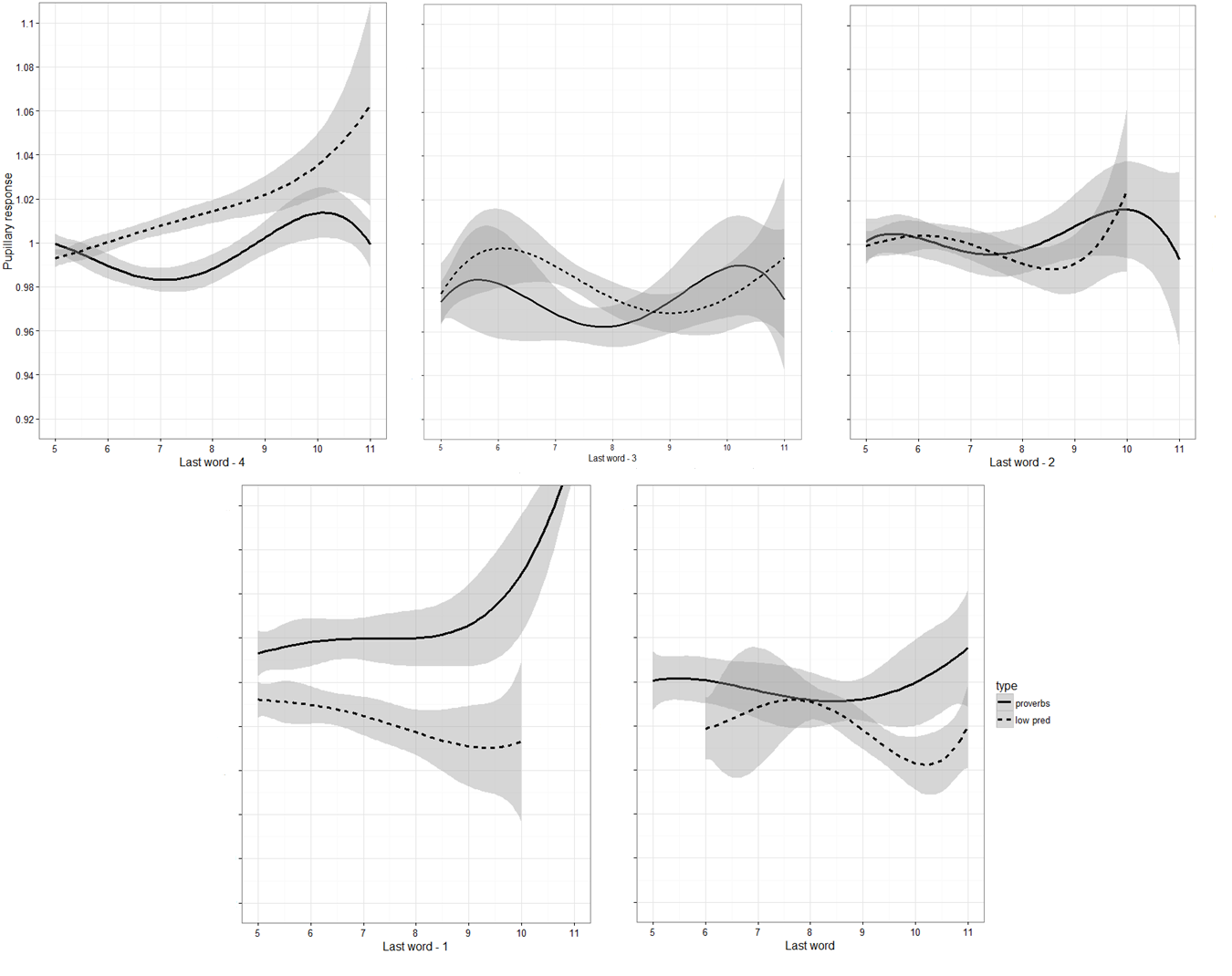}
\caption{Effect of word position in sentences on normalized pupil dilatation, broken down by regular sentences and proverbs. Panel reflects regression of normalized pupil dilatation on word on respective word position in the sentence. Shaded areas are $95\%$ confidence intervals.}
\label{fig:wp-sentence}
\end{figure}

When analyzing predictor per sentence type interaction (i.e., regular sentences vs. proverbs) we noted a significant and interesting effect of word number on pupil size ($t=-2.46$): regular sentences decreased readers' pupil size when comparing with proverbs. 
Proverbs increased pupil size from the middle to the end of the sentences (See Table \ref{tab:estimates} and Figure \ref{fig:nw}). 
Given that different proverbs have different word lengths and because the pupil increase tends to happen from the middle of the sentence we plotted a Figure that counts word position from the end of the sentence. 
In this way, the rightmost tick mark on the axis will always refer to the last word in the sentence/proverb, and the tick mark to its left refer to the second-to-last word (i.e., last word -1), and so on. 
Thus, when analyzing last word processing we appreciated that both sentences produced a similar effect on pupil dilatation although proverbs marginally increased pupil dilatation. 
Strikingly, the effect is different when analyzing \textit{Last word -1}, where proverbs clearly increased pupil dilatation. 
It seems that on these words readers recognize proverbs producing a memory cue with an impact on pupil dilatation (see Discussion for an explanation about the ``Eureka moment''). 
\textit{Last word -2} and \textit{word -3} showed similar pupil behavior when processing words, although \textit{Last word -3} in regular sentences evidenced a small increase on pupil dilatation when comparing with proverbs. 
On the other hand, \textit{Last word -4} showed that regular sentences produced a stronger increase on pupil dilatation. 
It seems that in \textit{Last word -4} proverbs are not yet identified and words in regular sentences produced a cognitive load increasing readers' pupil size.
Finally, only word predictability increased significantly pupil size where regular sentences showed an increased pupil when comparing with proverbs ($t=3.03$) (See Table \ref{tab:estimates} and Figure \ref{fig:pred}). 

\section{Discussion}
This work is, as far as we know, the first one analyzing pupil behavior during reading well defined words embedded in sentences with different contextual predictability. 
We measure pupil response of participants as they read regular sentences and proverbs. 
Using sentences with different contextual predictability allowed us to check whether readers devoted greater cognitive effort to process word information and to use contextual hints for improving their reading performance.

In the last 50 years, a large body of studies has confirmed that the pupil behavior changes depending of the task difficulty \citep{beatty1966pupillary, bradshaw1968pupil, hyona1995pupil}. 
In general, less effort decreased mean pupil dilatation. 
This common proxy between a reduced effort and a decreased pupil dilatation seems to be the pattern when analyzing longer words and low-predictable words (see Figures \ref{fig:length}, \ref{fig:pred}). 
In general, when more effort is required for processing words there is an increase in the mean pupil dilatation. 
Interestingly, when analyzing regular sentences vs. proverbs emerges a particular pupil behavior. 
It seems that readers' pupils increase when attentional cues (e.g, knowed words) are provided by proverbs (see Figure \ref{fig:nw}). 
As in \citet{papesh2012memory}, the pupil behavior was sensitive to the content of memory: sentences in which contextual predictability was stronger yielded the largest pupil dilatation. 
In accordance with \citet{gazzaley2013top}, predictive cueing resulted in improved working memory performance and it seems that this phenomenon increase pupil dilatation. 
\citet{just1993intensity} proposed that pupillary response as an indicator of how intensely the processing system is operating and interpreted the effect in terms of memory load. 
Our results suggest that in both, regular sentences and proverbs, the memory load is present when analyzing readers' pupil behavior (see Figures \ref{fig:length}, \ref{fig:pred} and \ref{fig:freq}). Additionally, this memory effect seems to be present from the middle of the sentences when analyzing proverbs and on the last words when looking at regular sentences (see Figure \ref{fig:nw} and Table \ref{tab:estimates}). 
\citet{schluroff1982pupil}, considered pupil size changes over time as a reflection of the difficulty of processing resulting from internal properties of the sentences. 
Our results suggest that both cognitive load in first place and task facilitation when retrieving upcoming words in the second one, increase pupil dilatation. 
Using regular sentences and proverbs allowed us to
explicitly vary the dynamics of memory processing during reading, because highly predictable words
in proverbs active top-down processes for predicting upcoming words (See \citet{fernandez2014eye} for sentences' description) words.

\citet{alnaes2014pupil} propose that cognitive factors, such as prior knowledge and expectations, and top-down attentional control interact with incoming sensory signals. They also affirm that this interaction may produce a bias in the competition between objects for access to the working memory \citep{corbetta2002control, desimone1995neural}.
The researchers propose that pupillary responses reflect the intensity of mental operations and the allocation of attention across a range of different tasks.
\citet{alnaes2014pupil} observed pupil associated activity in the Locus Coeruleus (LC). The LC is a small brain stem nucleus located in the rostral pons.  The noradrenergic projections of the LC are sent to virtually all brain regions with major density to areas known to be important in attentional processing. Further, pupil dilatation associated with cognitive processing are thought to result from an inhibitory effect on parasympathic oculomotor complex generated from the LC \citep{alnaes2014pupil}. 
Furthermore, the LC -norepinephrine (NE) based modulatory mechanism can help to establish attentional shifts of either, external or internal stimuli where one event becomes more relevant than each other \citep{laeng2012pupillometry}.
The LC is also engaged during the process of memory retrieval \citep{eschenko2008learning}. \citet{posner2008attention} have distinguished between alerting, orienting, and executive networks of the brain. 
In their model, the alerting network is innervated by the NE system and includes the LC, right frontal cortex, and regions of the parietal cortex. 
Thus, NE plays a crucial role in energizing the cortical system and promoting adequate levels of activation for cognitive performance. 
Similar to \citet{papesh2012memory}, our results suggest that when subjects need to allocate more resources for integrating current words and for predicting upcoming words i.e., when the Eureka moment emerge in proverbs \citet{fernandez2014eye} (see Figures \ref{fig:nw} and \ref{fig:wp-sentence}) an increase in their pupil sizes happens. 
As in previous works \citep{fernandez2014eye, fernandez2014lack, fernandez2015diagnosis} it seems that an upcoming predictability effect in proverbs increase pupil diameter while it is retrieved from memory (see Figure \ref{fig:wp-sentence}). 
As reported \citep{laeng2012pupillometry}, the LC response occurs about 100 ms after a relevant event and it takes an additional 60-70 ms for the activity within the LC to reach frontal cortex. Such a delay from the triggering event to NE release  at a cortical site is then 150-200 ms. 
This evidence suggest that pupillary responses could provide a signal of the moment in which the Eureka word emerges. 
Probing online comprehension processes during both regular sentences and proverbs and tracing their effects on pupil dilatation might give us a tool for the evaluation of cognitive effort and reading facilitation.
Our work replicate \citet{hyona2000morph} results (see Figure \ref{fig:length}), where longer words strongly increased pupil size irrespective of which kind of sentences we were analyzing. 
Finally, word frequency and word predictability influenced pupil responses. 
Previous researchers have documented that word frequency typically affects pupillary reflexes \citep{kuchinke2007pupillary, papesh2008pupil}. 
As with word frequency, our results show that more predictable words decreased pupil diameter.
Interestingly, the predictability effect was stronger for proverbs (see Figure \ref{fig:pred} and Table \ref{tab:estimates}).

To conclude, the present findings suggest that both regular sentences and proverbs increases and decreases pupil size.
Our work suggests that proverbs are retained in memory, aiding subsequent perception and recognition. 
As indicated by pupillometry, proverbs strongly influence subjective feelings of memory strength and cognitive demand. Our results show that pupil size dynamics may be a reliable measure to investigate the cognitive processes involved in sentence processing.

\section*{Acknowledgments}
We thank reviewer's comments that allowed us to  improve our paper. 
This work was partially supported by grants PGI 24/K062, Universidad Nacional del Sur and PICT 2013 - 0403, Agencia Nacional de Ciencia y Tecnología, Ministerio de Ciencia y Tecnología, Argentina.\\[5pt]

\noindent Preprint of an article published in Journal of Integrative Neuroscience, Online Ready, pp. 1-12, 2017. 

\noindent doi: 10.1142/S0219635216500266 

\noindent \textcopyright copyright World Scientific Publishing Company 

\noindent http://www.worldscientific.com/worldscinet/jin

\bibliographystyle{plainnat}
\bibliography{bibliography}

\end{document}